\documentclass[12pt]{iopart}

\usepackage{epsfig}

\begin{document}

\title{Scale--dependent Galaxy Bias}

\author{Peter Coles$^1$ and Pirin Erdo\u{g}du$^2$}
\address{$^1$ School of Physics \& Astronomy, Cardiff University,
Queens Buildings, The Parade, Cardiff CF24 3AA, United Kingdom}
\address{$^2$ School of Physics \& Astronomy, University of Nottingham,
University Park, Nottingham NG7 2RD, United Kingdom}

\ead{Peter.Coles@astro.cf.ac.uk}

\begin{abstract}
We present a simple heuristic model to demonstrate how feedback
related to the galaxy formation process can result in a
scale-dependent bias of mass versus light, even on very large
scales. The model invokes the idea that galaxies form initially in
locations determined by the local density field, but the subsequent
formation of galaxies is also influenced by the presence of nearby
galaxies that have already formed. The form of bias that results
possesses some features that are usually described in terms of
stochastic effects, but our model is entirely deterministic once the
density field is specified. Features in the large-scale galaxy power
spectrum (such as wiggles that might in an extreme case mimic the
effect of baryons on the primordial transfer function) could, at
least in principle, arise from spatial modulations of the galaxy
formation process that arise naturally in our model. We also show
how this fully deterministic model gives rise to apparently
stochasticity in the galaxy distribution.
\end{abstract}

\section{Introduction}

Thanks to large-scale spectroscopic  surveys such as the
Anglo-Australian 2dF Galaxy Redshift Survey (2dFGRS: Norberg et al.
2001; Wild et al. 2004; Conway et al. 2005) and the Sloan Digital
Sky Survey (Zehavi et al. 2002; Tegmark et al. 2004; Swanson et al.
2007) it is now well established that the clustering of galaxies
depends subtly on their internal properties. Since galaxies of
different types display different spatial distributions it follows
that not all galaxies can trace the distribution of underlying dark
matter. In other words galaxies are biased tracers of the
cosmological mass distribution. Theories of cosmological structure
formation must explain the relationship between galaxies and the
distribution of gravitating matter which probably yields important
clues to the process by which they were assembled.

Galaxy formation involves complex hydrodynamical and radiative
processes alongside the merging and disruption of dark matter
haloes. This entails a huge range of physical scales that poses
extreme  challenges even for the largest supercomputers. The usual
approach is therefore to encode the non-gravitational physics into a
series of simplified rules to be incorporated in a code which
evolves the dark matter distribution according to Newtonian gravity
(e.g. Benson et al. 2000). This ``semi-analytic'' approach has many
strengths, including the ability to make detailed models for direct
testing against observations, but it difficult to use it to make
models with which one can make inferences from data. For this
reason, simplified analytical models of bias are still extremely
useful if one hopes to proceed from observations to theory rather
than vice-versa.

In the new era of ``precision cosmology'' the presence of bias is
more an obstacle than a key to understanding (Zheng \& Weinberg
2007). Attempts to infer parameter values from cosmological
observations are hampered by the unknown relationship between
visible objects and the underlying mass fluctuations they trace. For
example, the relatively weak residual baryon acoustic oscillations
(BAO) one expects to be present in the matter power spectrum (Pen
1998; Meiksin, White \& Peacock 1999; Blake \& Glazebrook 2003;
Eisenstein et al. 2005; Seo \& Eisenstein 2005; Wang 2006) are
potentially extremely important diagnostics of the presence of dark
energy if they can be observed at high redshift. However, when
matter fluctuations are inferred from galaxy statistics, the form
and evolution of bias must be understood and controlled if the
required level of accuracy is to be reached. Here again simplified
anaytical models have an important role to play.

In this paper we introduce a simple yet general theoretical model
which can describe various aspects of galaxy bias is a unified way.
We describe biasing models in general in the next section. In
Section 3 we present our model and in Sections 4 and 5 we describe a
couple of applications. We discuss the results in Section 6.

\section{From Local Bias to the Halo Model}

The idea that galaxy formation might be biased goes back to the
realization by Kaiser (1984) that the reason Abell clusters display
stronger correlations than galaxies at a given separation is that
these objects are selected to be particularly dense concentrations
of matter. As such, they are very rare events, occurring in the tail
of the distribution function of density fluctuations. Under such
conditions a ``high-peak'' bias prevails: rare high peaks are much
more strongly clustered than more typical fluctuations (Bardeen et
al. 1986). More generally, in {\em local bias} models, the
propensity of a galaxy to form at a point where the total (local)
density of matter is $\rho$ is taken to be some function $f(\rho)$
(Coles 1993; Fry \& Gaztanaga 1993).

It is possible to place stringent constraints on the effect this
kind of bias can have on galaxy clustering statistics without making
any particular assumption about the form of $f$. In particular, it
can be shown that the large-scale two--point correlation function of
galaxies typically tends to a constant multiple of the mass
autocorrelation function in these models. Coles (1993) proved that
under weak conditions on the form of $f(\rho)$ as discussed in the
introduction, the large-scale biased correlation function of
galaxies would generally have a leading-order term proportional to
$\xi_{\rm m}(r)$. In other words, one cannot change the large-scale
slope of the correlation function of locally-biased galaxies with
respect to that of the mass. This was a serious problem for the
standard cold dark matter model of times past (which had
$\Omega_0=1$ and $\Lambda=0$) because there is insufficient power in
the matter spectrum in this model to match observations unless one
incorporates a strongly scale dependent bias (Bower et al. 1993).

The local bias ``theorem'' was initially proved for biasing applied
to Gaussian fluctuations only and did not necessary apply to galaxy
clustering where, even on large scales, deviations from Gaussian
behaviour are significant. Steps towards the plugging of this gap
began with Fry \& Gaztanaga (1993) who used an expansion of $f$ in
powers of the dimensionless density contract $\delta$ and weakly
non-linear (perturbative) calculations of $\xi_{\rm m}(r)$ to
explore the statistical consequences of biasing in more realistic
(i.e. non-Gaussian) fields. Based largely on these arguments,
Scherrer \& Weinberg (1998) showed explicitly that non-linear
evolution always guarantees the existence of a linear leading-order
term regardless of the form of $f$, thus strengthening the original
argument of Coles (1993) at the same time as confirming the validity
of the theorem in the non-linear regime. A similar result holds
under the hierarchical ansatz, as discussed by Coles et al. (1999).

It is worth noting that the original form of the local bias theorem
has a minor loophole: for certain peculiar forms of $f$ the leading
order term is proportional to $[\xi_{\rm m} (r)]^2$ (Coles 1993).
However, $\xi_{\rm m} (r)$ must be a convex function of $r$ because
its Fourier transform, the power spectrum, is non-negative definite
(i.e. it can be positive or exactly zero). Higher order terms  in
$\xi_{\rm m}^n$ therefore fall off more sharply than $\xi_{\rm }(r)$
on large scales so this loophole does not have any serious practical
consequences for large-scale structure.

Such results greatly simplify attempts to determine cosmological
parameters using galaxy clustering surveys, as well as facilitating
the interpretation of any specific features in large-scale
clustering statistics because they require the galaxy spectrum to
have the same shape as the underlying mass spectrum. This reduces
the possible effect of bias to a single parameter which can be
estimated and removed by marginalisation. On the other hand, it
results in a drastic truncation of the level of complexity in the
assumed relationship between galaxies and dark matter.

In hierarchical models, galaxy formation involves the formation of a
dark matter halo, the settling of gas into the halo potential, and
the cooling and fragmentation of this gas into stars. This all
happens within a population of haloes which is undergoing continuous
merging and disruption. Rather than attempting to model these stages
in one go by a simple function $f$ of the underlying density field
it is better to study the dependence of the resulting statistical
properties on the various ingredients of this process. Bardeen et
al. (1986), following Kaiser (1984), pioneered this approach by
calculating detailed statistical properties of high-density regions
in Gaussian fluctuations fields. Mo \& White (1996) and Mo et al.
(1997) went further along this road by using an extension of the
Press-Shechter (1974) theory to calculate the correlation bias of
halos, this making an attempt to correct for the dynamical evolution
absent in the Bardeen et al. approach. The extended Press-Schechter
theory forms the basis of many models for halo bias in the
subsequent literature (e.g. Matarrese et al. 1997; Moscardini et al.
1998; Tegmark \& Peebles 1998).

It is worth stressing that by  ``local bias'' we mean some form of
coarse--graining to select objects on a galaxy scale. In the earlier
models described above, galaxy correlations arise because the
underlying matter field is correlated but the process of galaxy
formation does not itself influence the formation of structure on
scales larger than this resolution scale. More recent developments
involve the Halo Model (Seljak 2000; Peacock \& Smith 2000; Cooray
\& Sheth 2002; Neyrinck \& Hamilton 2005; Blanton et al. 2006;
Schulz \& White 2006; Smith, Scoccimarro \& Sheth 2006, 2007). This
model generally assumes that galaxy properties are derived from the
underlying mass or halo field. Some degree of scale--dependence then
arises because galaxies interact on the scale of an individual halo
to provide some degree of self-organisation within the resolution
scale. This model has scored some notable successes at explaining
features in observed galaxy correlations.

It has also been suggested that bias might not be a deterministic
function of $\rho$, and that consequently there is a stochastic
element in the relationship between mass and light (Dekel \& Lahav
1999).

In the following sections we present a model that extends a number
of these different lines of thought. In particular we consider the
possibility that large-scale interactions between galaxies or
proto-galaxies might induce a significant scale dependent bias that
is qualitatively different from that which arises even in the halo
model.

\section{Self-interacting Galaxy Formation}

As described in the previous section, the idea of local bias models
is that the density of matter at a given spatial position ${\bf x}$
is responsible for generating the propensity that a galaxy will form
there (after suitable coarse-graining of the density field). In its
simplest terms we can represent this idea in terms of a galaxy
fluctuation field
\begin{equation}
\delta_{\rm g} ({\bf x})\equiv \frac{n({\bf x})}{\bar{n}}-1,
\end{equation}
where $n({\bf x})$ is the number density of galaxies at ${\bf x}$
and $\bar{n}$ is the mean number density of galaxies.
 The simplest way to account for
discreteness is to use the Poisson cluster model of Layzer (1956) in
which galaxies form with a probability proportional to $\delta_{\rm
g}$. If there are interactions within the resolution scale then the
Poisson model does not necessarily hold (Coles 1993). In order  to
keep the presentation of our model as simple as possible we ignore
discreteness effects and restrict ourselves to large scale
clustering properties. In local bias theories the galaxy field is a
deterministic function of the local matter density field at the same
point ${\bf x}$.

Our model for scale--dependent bias has the form:
\begin{equation}
\delta_{\rm g}({\bf x})  =  \delta_{\rm s}({\bf x}) + \alpha \int h
({\bf x-x'}) \delta_{\rm g} ({\bf x'}) d{\bf x'}
\end{equation}
In this equation the field $\delta_{\rm s}({\bf x})$ represents a
``seed'' field and the second term models the interactions. In a
realistic situation the parameter $\alpha$ might well be stochastic,
varying in a complicated way from galaxy to galaxy, but for
simplicity we will assume it to be a constant in this paper. In
principle a galaxy may either enhance or suppress the formation of
others around it so $\alpha$ may be either positive or negative. In
the absence of interactions (i.e. taking $\alpha=0$), the model
reduces to a standard biasing picture where the clustering of
galaxies is, at some level, reducible directly to the clustering ot
the mass. In the ``no-bias'' case the seed field will simply be the
underlying density fluctuation field, i.e. $\delta_{\rm
s}=\delta_{\rm m}$. Galaxies could then form as a Poisson sampling
of the mass field as suggested by Layzer (1956). For linear bias
models, we would take $\delta_{\rm s}=b\delta_{\rm m}$. In such
cases the resulting galaxy spectrum $P_{\rm g}(k)=b^{2} P_{\rm
m}(k)$ for all $k$. In general local bias models we might take the
seed field to be some local function $f(\delta_{\rm m})$, as
described in the previous section. In these cases $P_{\rm
g}(k)\simeq b^{2} P_{\rm m}(k)$ for small $k$ via the local bias
theorems. More realistically perhaps, $\delta_{\rm s}$ could be the
``halo field''. Explicitly in this case, and indeed implicitly in
the other cases discussed above, $\delta_{\rm s}$ does possess a
filtering scale of its own, with the width of the smoothing kernel
representing the characteristic size of a galaxy halo.

If the seed field is simply the halo field, the galaxies do not form
a Poisson sample; the distribution of galaxies within a given halo
is a degree of freedom within the halo model which must be fixed by
reference to observations (Seljak 2000; Peacock \& Smith 2000;
Cooray \& Sheth 2002). The seed field might also include stochastic
terms (Dekel \& Lahav 1999; Blanton et al. 1999; Matsubara 1999),
i.e. terms which can not be expressed as any function of $\rho_{\rm
m}$ but which might instead be modelled as random variables. The
first term on the right hand side of equation (2) therefore includes
the traditional bias models discussed in the previous section. If
$\alpha=0$ we recover models in which the clustering of galaxies is,
at some level, reducible directly to the clustering of the mass. In
such cases if the seed field were uncorrelated then all these models
would produce uncorrelated galaxies.

If $\alpha=0$ and the seed field is uncorrelated then all these
models would produce uncorrelated galaxies. If $\alpha\neq 0$,
however, then we have a qualitatively different form of bias. The
galaxy field then not only depends on the seed field, but also on
the galaxy field itself. This ``bootstrap'' effect allows a greater
degree of flexibility in modelling galaxy correlations. In
particular, even if the seed field were completely uncorrelated,
interactions could produce a non-zero galaxy-galaxy correlation
function in the bootstrap model. This can not happen in local bias
models. In this respect our model is similar to the autoregressive
(AR) models used to simulate time series: these are correlated
processes that are seeded by random (uncorrelated) noise. More
relevantly for cosmology, as we shall see shortly, the bootstrap
model allows us to generate scale-dependent bias that violates the
theorems referred to in Section 2. The initial seed field
$\delta_{s}({\bf x})$ plays the same role as the ``innovation'' in
autoregressive time series models.

The presence of the kernel  in equation (2) gives the model the
ability to generate non-local interactions if it extends over a
relatively large scale. The kernel $h({\bf y})$ determines the size
of the zone of influence of one galaxy on the formation of others in
its neighbourhood; we denote this scale by $R_{h}$. Just as with the
parameter $\alpha$, we take this scale to be constant for
simplicity. Note, however, that since both the scale  and level of
feedback may be difficult to predict given only the ambient density
field, it may be more realistic to model the kernel scale as
stochastic variable.

The filter should be defined in such a way that it preserves the
statistical homogeneity of the density field and does not lead to
diverging moments. For sensible filters $h$ will have the following
properties: $h=~{\rm constant}\simeq R_{h}^{-3}$ if $\vert {\bf
x}-{\bf x'} \vert \ll R_{h}$, $h \simeq 0$ if
 $\vert {\bf x}-{\bf x'} \vert \gg  R_{h}$,
$\int h({\bf y};R_{h}) d{\bf y} = 1$. We discuss a couple of
specific examples in the subsequent sections of this paper.

The integral on the right hand side of equation (2) represents the
galaxy fluctuation field convolved with a low pass filter. One can
write (2) in the form
\begin{equation}
\delta_{\rm g}({\bf x}) = \delta_{\rm s}({\bf x}) + \alpha
\delta_{\rm g} ({\bf x};
 R_{\rm h}).
\end{equation}
The filtered field, $\delta_g({\bf x}; R_{h})$,
 may be obtained by convolution of the ``raw'' galaxy density field with
some function $h$ having a characteristic scale $R_{h}$:
\begin{equation}
 \delta_{\rm g} ({\bf x}; R_{h}) = \int \delta_{\rm g}( {\bf x'}) h(
\vert {\bf x}- {\bf x'}\vert; R_{h}) d {\bf x'}. \end{equation} To
recover the local bias model with $\alpha \neq 0$ we simply take
$h({\bf x}-{\bf x'})=\delta_D({\bf x}-{\bf x'})$ in which case
$\delta_{\rm g}=\delta_{\rm s}/(1-\alpha)=b\delta_{\rm s}$. Scale
independence and linearity of the bias are therefore both recovered
in this limit.

Equation (2) is a Fredholm integral equation of the second type.
Assuming that the interaction kernel $h$ is well-behaved we can
solve it quite straightforwardly. Defining the Fourier transform of
$\delta_{\rm s}({\bf x})$ to be $\tilde{\delta}_{\rm m}({\bf k})$
etc and using the convolution theorem, the $k$-space version of the
equation (2) is seen to be
\begin{equation}
\tilde{\delta}_{\rm g}({\bf k})=\tilde{\delta}_{\rm s}({\bf k})+
\alpha \tilde{h} ({\bf k})\tilde{\delta}_{\rm g}({\bf k}),
\end{equation}
which gives a solution for $\tilde{\delta}_{\rm g}({\bf k})$:
\begin{equation}
\tilde{\delta}_{\rm g}({\bf k})=\frac{\tilde{\delta}_{\rm s}({\bf
k})}{1-\alpha \tilde{h} ({\bf k})}.
\end{equation}
The power spectrum of the filtered field is given by
\begin{equation}
P(k; R_{h}) = \tilde{h}^2 (k; R_{h}) P_{\rm g}(k),
\end{equation} where $P_{\rm g}(k)$ is the power spectrum of the
galaxy field. Assuming that $h({\bf y})$ is isotropic, the
galaxy-galaxy power spectrum can be expressed as
\begin{equation}
P_{\rm g}(k)= \frac{P_{\rm s}(k)}{|1-\alpha \tilde{h}(k)|^2},
\end{equation}
where $k=|{\bf k}|$. It is clear that the kernel can imprint
features into the power spectrum through the dependence on
$\tilde{h}(k)$, even in the case where $P_{\rm s}(k)$ is completely
flat. This means it is considerably more general than the simpler
models discussed above. It possesses some features that resemble the
cooperative galaxy formation model of Bower et al. (1993) but with
significantly more generality. We shall illustrate some of its
properties in the following sections.

\section{Bogus Baryon Wiggles?}

In this section we present an extreme example of scale--dependent
bias which is based on the idea that some violent astrophysical
process connected with galaxy formation (such as the ionizing
radiation produced by quasar activity) could seriously influence the
propensiy of galaxies to form in the neighbourhood of a given
object.  This concept is not new (Rees 1988; Babul \& White 1991),
and has been recently revived in a milder form (Pritchard,
Furlanetto \& Kamionkowski 2006).

To give an illustration of the extreme effects that could arise in
the galaxy power spectrum, consider the extreme example where the
zone of influence of a galaxy (or quasar) has a sharp edge similar
to an HII region. We can use our model to describe this situation if
we adopt a kernel which has the form of a {\it ``top hat'} filter,
with a sharp cut off, defined by the relation
\begin{equation}
 h_{\rm T}( \vert {\bf x}-{\bf x'} \vert; R_{\rm h})  = {3\over 4 \pi
R_{h}^3 } \Theta \Bigl( 1 -  {\vert {\bf x}-{\bf x'} \vert \over
R_{h}}\Bigr), \end{equation}
 where $\Theta$ is the Heaviside step
function: $\Theta(y)=0$ for $y\leq 0$ and $\Theta(y)=1$ for $y>0$.
The form of the kernel in Fourier space is then
\begin{equation} \tilde{h}_{\rm T}(k; R_{h}) = {3(\sin kR_{h} -
kR_{h}  \cos k R_{h}) \over (kR_{h})^3}~.
\end{equation}
Oscillatory features can be generated in the  galaxy power spectrum
by this form of interaction and with a suitable choice of scale
$R_h$ they could even mimic the BAOs mentioned in the Introduction.

\begin{figure}[htbp]
\centering \epsfxsize=\textwidth \epsfbox{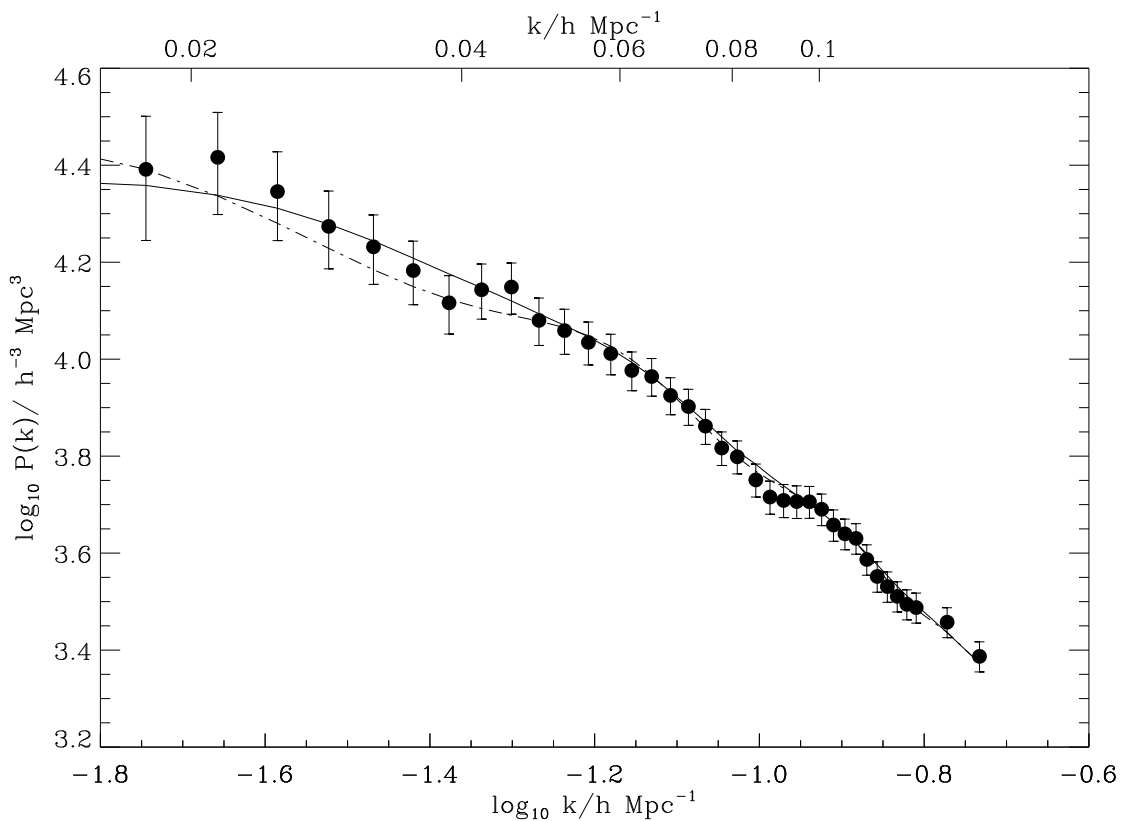} \caption{The
black filled circles and the associated error bars are the
  2dFGRS power-spectrum data given in Table 2 of Cole et al. (2005). The black
  solid lines in both plots denote
the reference power spectrum convolved with the
  window function also given in
  Table 2, with $\Omega_{\rm m}h=0.168$,
$\Omega_{\rm b}/\Omega_{\rm m}=0.17$ and $\sigma^{\rm gal}_8=0.89$.
The dashed
  lines are for the best fit model with biasing but no baryons. }
\label{fig1}
\end{figure}

\begin{figure}[htbp]
\centering \epsfxsize=\textwidth \epsfbox{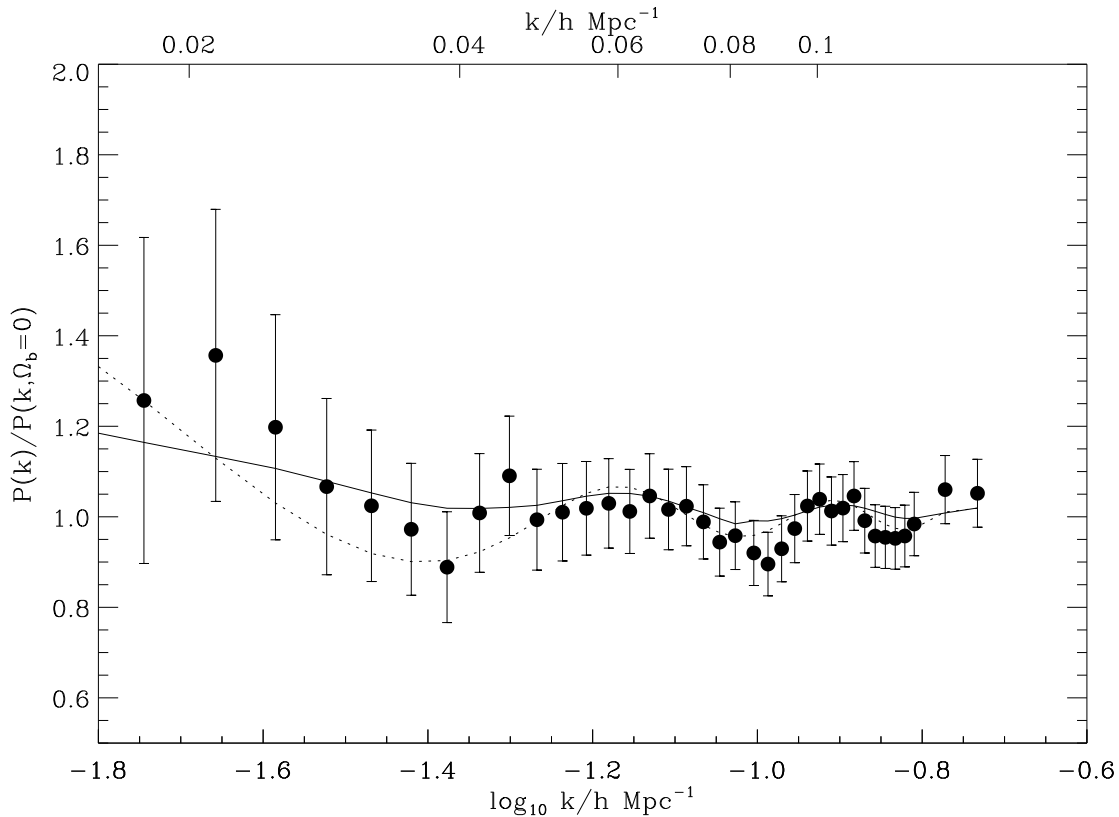} \caption{As
Figure 1, but with the data and the curves divided
  by a model with $b=1$, $\Omega_{\rm m}=0.23$ and $\Omega_{\rm b}=0$.} \label{fig2}
\end{figure}

To establish the required parameters we refer to the 2dFGRS
redshift-space power spectrum data given in Table 2 of Cole et al.
(2005) for the 2dFGRS. We do not attempt to fit the small-scale
clustering in this data set. This could be done by fiddling with the
form of $\delta_{\rm s}$, but our interest lies here in illustrating
the large--scale behaviour only. We also ignore redshift--space
distortions. In Cole et al. (2005), the error bars on the spectrum
are derived from the diagonal elements of the covariance matrix
calculated from model lognormal density fields. The model power
spectrum for these lognormal fields has $\Omega_{\rm m}h=0.168$,
$\Omega_{\rm b}/\Omega_{\rm m}=0.17$ and $\sigma^{\rm g}_8=0.89$ and
agrees very well with the best fit model for the overall 2dFGRS
power spectrum. This model, convolved with the 2dFGRS survey window
function, is also given in Table 2 of Cole et al. (2005) and plotted
in Figure~\ref{fig1} (solid line) \& Figure~\ref{fig2}. Using the
full covariance matrix, Cole et al (2005) find $\chi^2/{\rm
d.o.f}=37/33$ for $k<0.2$  $h {\rm Mpc}^{-1}$. As this analysis is
for illustrative purposes only , we do not perform a full likelihood
analysis, rather we calculate the $\chi^2$ for the same model using
only the error bars. In this case the fit is characterized by
$\chi^2/{\rm d.o.f}=12/33$.  As discussed in Section 5 of Cole et al
(2005), since the convolution with the survey window function causes
the errors to be correlated, resulting in a very low value of
$\chi^2$. The goodness of fit does however provide a useful
benchmark for our alternative explanation of the wiggles seen in
$P(k)$.

In order to explain the shape of the galaxy spectrum using only
galaxy interactions and without the baryon oscillations, we use a
top-hat kernel for our biasing model and fit it to the same data
using the Eisenstein \& Hu (1998) transfer functions and assuming
$n_{\rm s}=1$, $h=0.72$ and $\Omega_{\rm
  b}=0$. In other words we use an underlying cosmology without baryon oscillations and seek to explain the shape of the
  galaxy spectrum using only galaxy interactions. Our best fit cosmological parameters are $\Omega_{\rm m}=0.23$,
$\sigma^{\rm g}_8=0.85$ and for the bias model we get $\alpha=0.25$
and $R_h=114 {\rm Mpc}$. This model has $\chi^2/{\rm d.o.f}=9/33$.
The value of $\chi^2$ is again very low due to correlations between
the data points,  but a comparison with the result of the previous
paragraph for
  which the same problem also holds, demonstrates that the fit is if
  anything marginally better for our model than for the reference model used by Cole et al. (2005).

Of course one does not know for sure whether and how ionization
influences galaxy formation, but this example illustrates that in
principle the observed wiggles in the galaxy power spectrum could
have an astrophysical rather than cosmological origin. This would
pose problems for their use as cosmological probes. On the other
hand, the scale required is very large. Rees (1988) pointed out that
a quasar of luminosity $L{\rm uv}$ lasting for a time $t_{\rm Q}$
produces sufficient energetic photons to ionize all the baryons
within a radius
\begin{equation}
R_h \simeq 67 \left( \frac{L_{\rm uv}}{10^{46} \, {\rm erg}\, {\rm
s}^{-1}} \right)^{1/3} \left(\frac{t_{\rm Q}}{2\times 10^9 \,{\rm
yrs}}\right)^{1/3} \,\mbox{Mpc}.
\end{equation}
In order to be able to contribute at a redshift $z$, the ionizing
photons must have been emitted in less than the lifetime of the
Universe at that redshift, $t(z)$. This places a minimal requirement
that $t_Q<t(z)$. In the concordance cosmology, $t(z=3)\simeq 2.2$
Gyrs, $t(z=6)\simeq 0.95$ Gyr and $t(z=10)=0.48$ Gyr. The actual
lifetime of quasars may well depend on their mass, but recent
estimates suggest $t_Q\simeq 10^8$ yrs is more likely than $10^9$
yrs (Mclure \& Dunlop 2004). If this is the case then equation (11)
implies that the corresponding value is more like  $R_h\simeq 25$
Mpc by equation (11); for this value of $t_Q$ the  required
ionization could easily have been achieved early, but the scale of
resulting wiggles would be relatively small. For $R\simeq 100$ Mpc
one needs to push the parameters excessively hard: a high value of
$t_Q>2\times 10^{9}$ and a redshift of reionization $z<3$ would be
necessary. This seems to be at odds with the general consensus that
reionization of the Universe happened relatively early (Becker et
al. 2001; Fan et al. 2002).

There are other problems with this model. Quasars have a range of
lifetimes and luminosities. Their radiation may also be beamed
rather than isotropic. And in any case it is not known to what
extent the galaxy formation process is sensitive to this form of
feedback anyway. Moreover, the baryon acoustic oscillations inferred
from galaxy clustering have the same characteristic scale as that
derived from cosmic microwave background observations. This would be
a sheer coincidence in our model.

This model is therefore unlikely to be the  correct interpretation
of  observed wiggles, but it does at least demonstrate that
large-scale interactions can have a significant impact on the shape
of the clustering power spectrum. Notice also that even if the scale
$R_h$ is not sufficiently large to match the observed oscillations,
any non-zero astrophysical effect could seriously degrade the
ability to recover cosmological information from galaxy surveys.
Mass tracers selected in some way other than counting galaxies may
well display clustering that is less susceptible to this type of
feedback bias. Galaxy clusters may be detected not only detected
through X-ray emission or Sunyaev-Zel'dovich measurements, both of
which are sensitive to the properties of the extremely hot gas the
clusters contain. If these properties vary systematically on large
scales then scale-dependent bias may also apply to such objects.
However, the strong non-linear merging and heating processes that
create this intracluster gas are likely to swamp any primordial
effects generated on smaller scales. One would therefore expect
cluster correlations to be less vulnerable to astrophysical
modulation than galaxy correlations; complementary observations on
the same length scales could be be used to identify and eliminate
this source of uncertainty.

\section{Scale-dependence versus Stochasticity}

Even if the scale and form of the interaction kernel do not produce
very large scale features in the galaxy correlation function or
power spectrum, it is still possible for scale--dependence to
manifest itself in more subtle ways. In particular, it is possible
for scale-dependence to appear as a form of stochastic bias (Dekel
\& Lahav 1999) even though the relationship (2) is entirely
deterministic once the density field is specified.

To see how this happens consider a simplified version of our general
model in which the seed field $\delta_s$ is simply the matter
density field $\delta_m$. Let us assume explicitly that the fields
were are considering are filtered on a scale $R_0$ to represent the
selection of galaxy sized objects. Let the scale of
feedback--induced interactions be $R_{\rm F}$, so that
\begin{equation}
\delta_{\rm m}(R_0)=\delta_{\rm g} (R_0) - \alpha \delta_{\rm g}
(R_{\rm F}).
\end{equation}
It is straightforward to see that
\begin{eqnarray}
\langle \delta_{\rm m} \delta_{\rm g} \rangle & = & \langle
\delta_{\rm g} (R_0)^2 \rangle - \alpha \langle \delta_{\rm g}
(R_{\rm F}) \delta_{\rm g} (R_0) \rangle \nonumber\\
 & = & \langle \delta_{\rm g}^2 \rangle \left( 1-\alpha
 \frac{\langle \delta_{\rm g}(R_{\rm F}) \delta_{\rm g}(R_0)\rangle}{\langle \delta_{\rm g} (R_0)^2 \rangle}\right)
\end{eqnarray}
and
\begin{eqnarray} \langle \delta_{\rm m}^2 \rangle & = &\langle
\delta_{\rm g} (R_0)^2 \rangle + \alpha^2 \langle \delta_{\rm
g}(R_{\rm F})^2\rangle - 2\alpha \langle \delta_{\rm g} (R_0)
\delta_{\rm g} (R_{\rm F}) \rangle\nonumber \\
&=& \langle \delta_{\rm g}^{2}\rangle \left\{ 1+ \alpha^2
\frac{\langle \delta_{\rm g}(R_{\rm F}) ^2\rangle}{\langle
\delta_{\rm g} (R_0)^2 \rangle} - 2 \alpha \frac{\langle \delta_{\rm
g} (R_0) \delta_{\rm g}(R_{\rm F}) \rangle}{\langle \delta_{\rm
g}(R_0)^2 \rangle} \right\},
\end{eqnarray}
where we have dropped the dependence on $R_0$ in the terms outside
the curly brackets. It is useful to define the quantities
\begin{equation}
\gamma \equiv \frac{\langle \delta_{\rm g} (R_{\rm F}) \delta_{\rm
g} (R_0) \rangle}{\langle \delta_{\rm g}(R_0)^2 \rangle}
\end{equation}
and \begin{equation} \omega^2 \equiv \frac{\langle \delta_{\rm g}
(R_{\rm F})^2 \rangle}{\langle \delta_{\rm g}^2 (R_0) \rangle},
\end{equation}so that the cross-correlation coefficient between the
mass and galaxy fluctuation fields is
\begin{equation}
r \equiv \frac{\langle \delta_{\rm m} \delta_{\rm g}
\rangle}{\langle \delta_{\rm g}^2 \rangle^{1/2} \langle \delta_{\rm
m}^2 \rangle^{1/2}} = \frac{1-\alpha \gamma}{(1+\alpha^2 \omega^2 -
2\alpha \gamma)^{1/2}}.
\end{equation}

To provide a simple illustrative model we assume a {\em Gaussian
filter}:
\begin{equation}
h_{\rm G}( \vert {\bf x}-{\bf x'} \vert; R_{\rm F}) = {1 \over (2
\pi R_{\rm F}^2)^{3/2}} \exp \Bigr (- {\vert {\bf x}- {\bf x'}\vert
^2 \over 2 R_{\rm F}^2}\Bigr), \end{equation} for which the
appropriate window function is
\begin{equation} \tilde{h}_{\rm G} (k R_{\rm F}) = \exp
\Bigl[-{(k R_{\rm F})^2 \over 2} \Bigr].
\end{equation}
We then need to tackle quantities of the form
\begin{equation}
\langle \delta_{\rm g} (R_1) \delta_{\rm g} (R_2) \rangle =
\frac{1}{2\pi^2} \int dk k^2 P_{\rm g} (k) \exp [ -
k^2(R_1^2+R_2^2)],
\end{equation}
which can be evaluated straightforwardly if we assume, for
simplicity,  that the (unsmoothed) galaxy power spectrum is a
power-law: $P_{\rm g}(k) \propto k^{n}$. In this case we find that
\begin{equation}
\langle \delta_{\rm g} (R_0) \delta_{\rm g} (R_{\rm F}) \rangle =
\sigma^2 \left( \frac{2R_0^2}{R_0^2 + R_{\rm F}^2}\right)^{n+3/2},
\end{equation}
where $\sigma^2$ is the variance of the unsmoothed density field.
This gives
\begin{equation}
\gamma = \left( \frac{2R_0^2}{R_0^2 + R_{\rm F}^2}\right)^{n+3/2}
\end{equation}
and
\begin{equation}
\omega^2 = \left( \frac{R_{0}}{R_{\rm F}} \right)^{(n+3)}.
\end{equation}
Note that if $R_0=R_{\rm F}$ so that the feedback scale is no larger
than a galaxy scale then $\omega=1$, $\gamma=1$ and consequently
$r=1$. If, however, $R_{\rm F}>R_0$ then $\gamma<1$. However, it is
always true that $\omega^2 > \gamma$ so that $(1-\alpha\gamma)^2< 1+
\alpha^2 \omega^2 - 2 \alpha\gamma$ and consequently that $r<1$. The
larger the value of $R_{\rm F}$ compared to $R_{0}$ the smaller the
resulting value of $r$.

Assuming the fields $\delta_{\rm m}$ and $\delta_{\rm g}$ are
jointly Gaussian one can express the conditional distribution of one
given a specific value of the other. Suppose the (unconditional)
variance of $\delta_{\rm g}$ is $\sigma^2$ then the variance after
conditioning on $\delta_{\rm m}=a$, say, reduces to
$\sigma^2(1-r^2)$. Only if $|r|=1$ is there no scatter in the
relationship. For this reason a value of $r<1$ is usually taken to
indicate the presence of stochastic bias (e.g. Tegmark \& Bromley
1999), but in this case the  scatter in the relationship between
$\delta_{\rm m}$ and $\delta_{\rm g}$ arises from non-locality in a
fully deterministic way. This suggests that considerable care needs
to be exercised in the interpretation of measured values of $r$ :
they may be indicative of scale--dependence rather than stochastic
effects.

If we instead look at the galaxy and matter fields (assuming
$\delta_s=\delta_m$) in Fourier space the situation is quite
different. In this case, by equation (8) we get
\begin{equation}
P_{\rm g}(k) = b^2(k) P_{\rm m}(k)
\end{equation}
with $b(k)=1-\alpha\tilde{h}(k)$. The cross-spectrum in Fourier
space is usually defined to be $P_{\rm mg}=r(k) b(k)P_{\rm m}$
(Tegmark \& Bromley 1998) for stochastic bias, with $r(k)$ playing a
role analogous to the correlation coefficient discussed above. In
this case, however, it reduces to $P_{\rm mg}= b(k)P_{\rm m}$
indicating a complete absence of stochasticity. The apparent
stochasticity in real space is actually due to non-locality, but the
model is local (and linear) in Fourier space so no stochasticity
appears in this representation. This is an example of a phenomenon
noted by Matsubara (1999).

\section{Discussion and Conclusions}

In this paper we have presented a new model for scale-dependent
astrophysical bias. Although it is inspired to some extent by Bower
et al. (1993), this model is considerably more general and easier to
use. In the absence of any more complete theory of galaxy formation
we hope it will provide a useful way to parametrise the possible
level and scale of interactions so that they can be determined from
observations and eliminated from cosmological considerations.

We illustrated the generality of this model by pushing it to an
extreme and showing that it can produce features that mimic baryon
oscillations. Although the required effect is quite small in
amplitude it does require astrophysical processes to be coordinated
over very large scales. This, together with the concordance between
clustering observations and the cosmic microwave background,
suggests that the observed wiggles have a primordial origin.
Nevertheless, in the precision era, any scale dependence in
clustering bias could seriously degrade the business of cosmological
parameter estimation. However, as we have argued in Section 4,
different forms of mass tracer are unlikely to suffer from this bias
to the same extent as galaxies. Using complementary observations
should provide sufficient data to estimate the parameters in our
bias model. This will not only allow us to learn whether there is
significant evidence for scale-dependent bias at all but also, by
marginalization, provide a way to remove this uncertainty from
cosmological studies. Some of the observations will go towards
estimating and eliminating a nuisance parameter rather than reducing
the statistical uncertainty in interesting ones so the existence of
scale-dependent bias will degrade the cosmological value of surveys
to some extent even if it can be modelled satisfactorily.

As a second, less extreme example of our approach we showed how
non--locality in the feedback relationship described by equation (2)
bears many of the hallmarks of stochastic bias. In particular,
although our model is deterministic once the density field is
specified, it is characterized by an imperfect correlation between
galaxy and mass fluctuations. The difference between our model and a
truly stochastic one is that in our case the residuals are not
random but correlated through the interaction terms. One might learn
more from observations by looking for correlated scatter than by
giving up and treating them as completely stochastic. In any case
the model we have presented shows up a terminological deficiency:
stochasticity and non-locality can be easily confused.

\section*{Acknowledgments}

We acknowledge support from PPARC grant PP/C501692/1.

\References

\item[] Babul A and White S D M 1991 {\it Mon. Not. R. astr. Soc.} {\bf 253} L31

\item[] Bardeen J M, Bond J R, Kaiser N and Szalay A S 1986 {\it Astrophys. J.} {\bf 304} 15

\item[] Becker R H et al 2001 {\it Astron. J.} {\bf 122} 2850

\item[] Benson A J, Cole S, Frenk C S, Baugh C M and  Lacey C G 2000
{\it Mon. Not. R. astr. Soc.} {\bf 311} 793

\item[] Blake C and Glazebrook K 2003 {\it Astrophys. J.} {\bf 594}
665

\item[] Blanton M R,  Cen R, Ostriker J P and Strauss M A 1999 {\it
Astrophys. J.} {\bf 522} 590

\item[] Blanton M R, Eisenstein D H, Hogg D W and Zehavi I 2006 {\it
Astrophys. J.} {\bf 645} 977

\item[] Bower R G, Coles P, Frenk C S and White S D M 1993 {\it Astrophys. J.} {\bf 405} 403

\item[] Cole S et al. 2005 {\it Mon. Not. R. astr. Soc.} {\bf 362}
505

\item[] Coles P 1993 {\it Mon. Not. R. astr. Soc.} {\bf 262} 1065

\item[] Coles P, Melott A L and Munshi D 1999 {\it Astrophys. J.} {\bf 521} L5

\item[] Conway E et al 2005 {\it Mon. Not. R. astr. Soc.} {\bf 356} 456

\item[] Cooray A and Sheth R 2002 {\it Phys. Rep.} {\bf 372} 1

\item[] Dekel A and Lahav O 1999 {\it Astrophys. J.} {\bf 520} 24

\item[] Eisenstein D J et al 2005 {\it Astrophys. J.} {\bf 633} 560

\item[] Eisenstein D J and Hu W., 1998, {\it Astrophys. J.}, {\bf 496} 605

\item[] Fan X et al. 2002 {\it Astron. J.} {\bf 123} 1247

\item[] Fry J N and Gaztanaga E 1993 {\it Astrophys. J.} {\bf 413}
447

\item[] Kaiser N 1984 {\it Astrophys. J.} {\bf 284} L9

\item[] Layzer D 1956 {\it Astrophys. J.} {\bf 61} 383

\item[] Matarrese S, Coles P, Lucchin F and Moscardini L 1997 {\it Mon. Not. R. astr. Soc.} {\bf 286} 115

\item[] Matsubara T 1999 {\it Astrophys. J.} {\bf 525} 543

\item[] McClure R J and Dunlop J S {\it Mon. Not. R. astr. Soc.}
{\bf 352} 1390

\item[] Meiksin A, White M and Peacock J A 1999 {\it Mon. Not. R.
astr. Soc.} {\bf 304} 851

\item[] Mo H and White S D M 1996 {\it Mon. Not. R. astr. Soc.} {\bf 282} 347

\item[] Mo H, Jing Y and White S D M 1997 {\it Mon. Not. R. astr. Soc.}
{\bf 284} 189

\item[] Moscardini L, Coles P, Lucchin F and Matarrese S 1998 {\it
Mon. Not. R. astr. Soc.} {\bf 299} 95

\item[] Neyrinck M C, Hamilton A J S and Gnedin N Y 2005 {\it Mon. Not. R. astr.
Soc.} {\bf 362} 337

\item[] Norberg P et al 2001 {\it Mon. Not. R. astr. Soc.} {\bf
328} 64

\item[] Peacock J A and Smith R E 2000 {\it Mon. Not. R. astr. Soc.}
{\bf 318} 1144

\item[] Pen U 1998 {\it Astrophys. J.} {\bf 504} 601

\item[] Press W H and Schechter P L 1974 {\it Astrophys. J.} {\bf 187} 425

\item[] Pritchard J R, Furlanetto S R and Kamionkowski M 2006
astro-ph/0604358

\item[] Rees M J 1988 in {Large Scale Structures of the Universe},
IAU Symposium No. 130, eds Audouze J, Pelletan M-C and Szalay A S.
Kluwer, Dordrecht

\item[] Scherrer R J and Weinberg D H 1998 {\it Astrophys. J.} {\bf 504} 607

\item[] Schulz A E and White M 2006 {\it Astroparticle Phys.} {\bf
25} 172

\item[] Seljak U 2000 {\it Mon. Not. R. astr. Soc.} {\bf 318} 2003

\item[] Seo H J and Eisenstein D J 2005 {\it Astrophys. J.} {\bf 633} 575

\item[] Smith R E, Scoccimarro R and Sheth R K 2006 astro-ph/0609547

\item[] Smith R E, Scoccimarro R and Sheth R K 2007 astro-ph/0703620

\item[] Swanson M E C, Tegmark M, Blanton M and Zehavi I 2007
astro-ph/0702584

\item[] Tegmark M and Bromley B C 1999 {\it Astrophys. J.} {\bf 518}
L69

\item[] Tegmark M and Peebles P J E 1998 {\it Astrophys. J.} {\bf 500} L79

\item[] Tegmark M et al 2004 {\it Astrophys. J.} {\bf 606} 702

\item[] Wang Y 2006 {\it Astrophys. J.} {\bf 647} 1

\item[] Wild V et al 2004 {\it Mon. Not. R. astr. Soc.} {\bf 356} 247

\item[] Zehavi I et al. 2002 {\it Astrophys. J.} {\bf 571} 172

\item[] Zheng Z and Weinberg D M 2007 {\it Astrophys. J.} {\bf 659}
1

\endrefs

\end{document}